\documentclass[12pt]{article}
\usepackage{amsfonts}
\usepackage[dvips]{graphicx}
\begin{document}

\title{"Paradoxical" coexistence of continuum of 'bound' waves and discrete set
of unlimited motion states}
\author{V.M. Chabanov , B.N. Zakhariev \\ Laboratory of
theoretical physics, Joint Institute for Nuclear Research,
\\ 141980 Dubna, Russia \\
E-mail addresses: zakharev@thsun1.jinr.ru;
chabanov@thsun1.jinr.ru}
\date{}

\maketitle

\begin{abstract}
We have combined two remarkable phenomena: resonance tunneling and
Anderson localization. It results in unexpected  spectrum reverse
to usual notions. It is demonstrated by the quantum system with
chaotic distribution of  potential resonance tunneling traps over
the whole coordinate axis. The corresponding spectrum  contains
continuum of 'bound' states (Anderson's localization) and discrete
tunneling resonances of unlimited wave propagation. It is in
contrast to the  usual situation with discrete bound and continuum
of scattering states.
\end{abstract}


\section{Introduction }
It is instructive to consider different simplifying limiting cases
and paradoxical combinations of properties of  quantum systems.
So, we can extract clear elementary constituents of complex
systems.   Our last book "Submissive quantum mechanics: new status
of the theory in inverse problem approach" \cite{obed} contains a
lot of such our model discoveries, see also [2-8]. In this paper
we demonstrate one more example of construction intriguing model
with 'inverse spectrum' of discrete states for which randomly
distributed potential obstacles are transparent for waves and
continuum states localized waves by Anderson's mechanism
\cite{Ander}. But at first we shall consider the resonance
tunneling phenomenon by the model of two $\delta$-spikes.
Afterwards, we shall construct a model of randomly spaced traps,
each is built as such a two-delta spikes system.

\section{Model of resonance tunneling through two delta-potential barriers}
Let us consider the simplified explanation of the resonance
tunneling phenomenon discovered long ago, but usually not clearly
understandable.
 Consider two identical barriers. A small part of wave flux
penetrating through the first barrier, occurs in inter-barrier
trap on quasi-bound energy level due to specially choice of the
incident wave energy. In this case inside the trap the multiple
reflected waves from the confining potential walls does not
suppress themselves due to constructive interference. It is
difficult to escape the trap and the waves accumulate inside it.
Finally the amplitude of waves between barriers increases so much
that the intensity of backward decaying waves becomes equal to the
reflected ones, but can have the opposite phase  that completely
destroys the reflection. The excellent illustration  of this can
serve the pictures in books by Brandt and Dahmen \cite{BD,BDS}.
 It is also shown in Fig 1.

 \begin{figure}[h]
\begin{center}
\includegraphics[height=8cm,width=11.8cm]{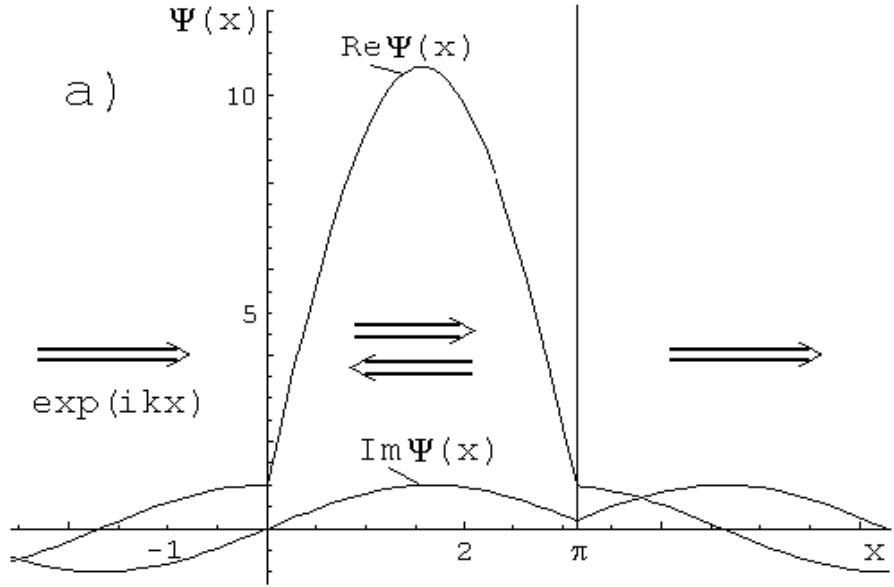}
\end{center}
\caption{{\small  The scattering waves (real and imaginary parts
of the Jost solution) at the energy  of resonance tunneling
through two $\delta$-barriers $10 \delta (x) + 10 \delta (x-\pi)$.
Note the wave accumulation between the potential barriers which
occurs here  for only one of two independent solutions. } }
\end{figure}

We can also construct the model solution for transparency of two
weakly penetrable barriers  in form of linear combination  of
simple solutions for  separate barriers. As the first solution we
shall take the incident wave  $\exp(ikx)$ on the potential
delta-barrier $V_{0}\delta (x)$ at the origin  $x=0$, see Fig.2.

 \begin{figure}[h]
\begin{center}
\includegraphics[height=7cm,width=6.8cm]{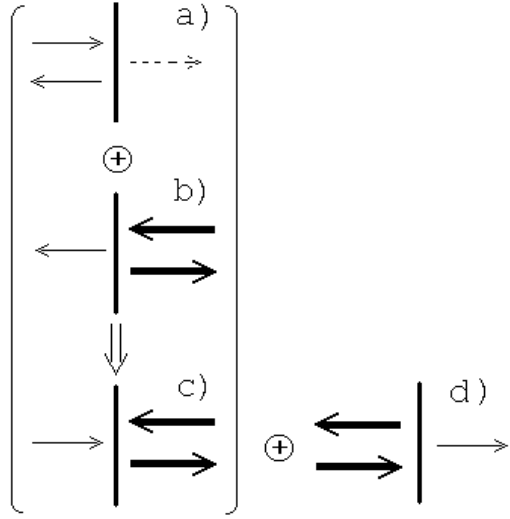}
\end{center}
\caption{{\small   Schemes of elementary tunneling  through one
barrier. Of these simplest processes ($a+b->c+d$) can be
constructed the solution for the system of two barriers with the
total (resonance) transparency. } }
\end{figure}

\noindent  The penetration and reflection coefficients
$T(k),\enskip
 R(k)$, the  factors before the  transit  and  reflected waves: a)
 $ T(k) \exp(i k x) $ at  $x > 0, \enskip k=\sqrt{e}$ and $ R(k) \exp(-i k x)
  + \exp(i k x)$ at $x <0$  have in this case the form
 \begin{eqnarray} T(k)=2ik/(2ik-V_{0})
;\,\,\, R(k)=V_{0}/(2ik-V_{0}).  \label{T} \end{eqnarray}

  \begin{eqnarray} \Psi
(x\le 0)= T(k) \exp(-ikx); \\ \Psi (x\ge 0)= \exp(-ikx)+ R(k)
\exp(ikx). \label{P}
\end{eqnarray}
We choose the normalization of this solution (we multiply this
solution (\ref{P}) by $-R/T$) so that the transit wave differs
only by the sign form the reflected wave of the first solution.
Let us now add the solutions (\ref{T}) and the renormalized one
(\ref{P}). As a result there is left only the wave moving to the
right while waves going to left cancel one another, see Fig.2b.The
obtained  solution, see Fig.2 c, can be smoothly joined with the
solution (\ref{T}) for the barrier shifted to the right, see Fig.2
d renormalized so that both these solutions can coincide between
the barriers. The necessary condition for  this is the coincidence
of penetration rates through the left and right barriers
separately at the energy of resonance tunneling. Then the absolute
values of wave amplitudes inside the region of inter-barrier
motion coincide for both solutions. The equality of phases can be
achieved by the choice of the relative distance between the
barriers at fixed energy or by variation of the energy of
solutions (which is feasible provided the penetrability for each
barrier is the same, i.e., the barriers are identical). As a
result we get the desired solution for total transparency.

\begin{figure}[h]
\begin{center}
\includegraphics[height=8cm,width=9.756cm]{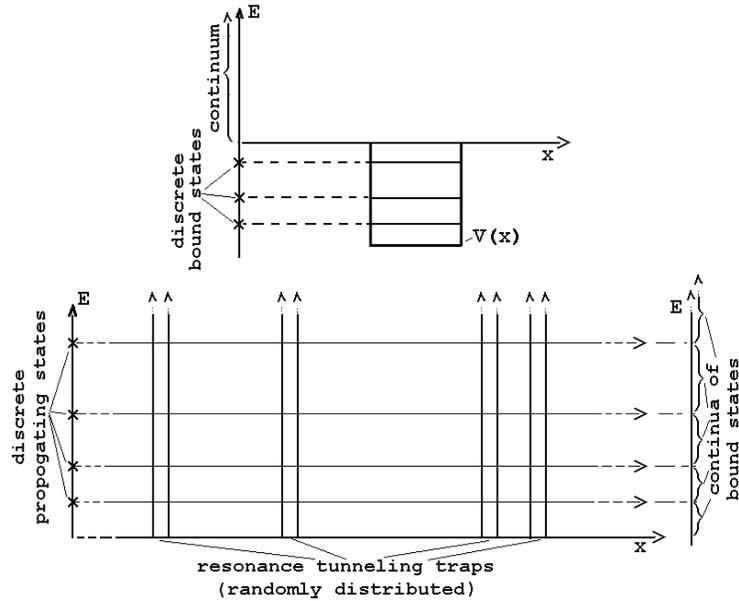}
\end{center}
\caption{{\small Upper part: usual  quantum spectrum with discrete
bound states in potential well and continuum of scattering states.
Lower part: "paradoxical"  spectrum of discrete propagating waves
and continua of Anderson's localized  (bound) states.} }
\end{figure}

\section{Wave motion in the system of infinite number of resonance
tunneling traps randomly distributed over the x-axis }

Let all these traps be identical the the considered above two
delta-barrier potentials. The whole complex of these traps will be
absolutely transparent at discrete resonance tunneling energy
values. So these waves can propagate along the whole x-axis from
$-\infty$ to $+\infty$  without reflection from the traps, see
Fig.3. But in the continuum of energy values between the tunneling
resonances there will be reflection of waves. So the random
distribution of the traps must result in Anderson localization of
all these waves. Thus, instead of usual situation when there is
discrete spectrum of bound states and continuum scattering states
we have here "paradoxically" inverse picture. These curious
possibility enriches our quantum intuition.

\section{Conclusions}
The real physical systems are usually too complicated  for
understanding. But recently revealed clear and exactly solvable
models [1-8] forming complete sets enter often as elementary
constituents into these objects. They are hidden there in the
mixture with multiple other details of their structure. But after
mastering the qualitative essence of these instructive models, the
analysis of the entangled picture, being previously beyond our
understanding, can become significantly simplified.
 Here we have considered
one more phenomenon which is, undoubtedly an additional step to
 comprehension of quantum world.

\end{document}